\documentclass[sigconf]{acmart}
\AtBeginDocument{%
  }

\setcopyright{acmlicensed}
\copyrightyear{2018}
\acmYear{2018}
\acmDOI{XXXXXXX.XXXXXXX}
\acmConference[Conference acronym 'XX]{Make sure to enter the correct
  conference title from your rights confirmation email}{June 03--05,
  2018}{Woodstock, NY}
  
\acmISBN{978-1-4503-XXXX-X/2018/06}

\usepackage{algorithm}
\usepackage{algpseudocode}
\usepackage{xcolor}
\usepackage{colortbl}

\usepackage{amsfonts} % for additional math symbols
\usepackage{amssymb}  % for extra symbols
\usepackage{mathtools} % for advanced mathematical tools
\usepackage{amsmath}
\usepackage{hyperref}
\DeclareMathOperator*{\argmax}{argmax}

\begin{document}

\title{Thinking Longer, Not Larger: Enhancing Software Engineering Agents via Scaling Test-Time Compute}

% Scaling Minds, Not Models: Test-Time Reasoning Enhancement for Resource-Efficient Software Engineering Agents

% TTS4SWE: A Unified Test-Time Scaling Framework for Deployable Software Engineering Agents

% 24GB to SOTA: Redefining SWE Agent Efficiency through Dual Test-Time Scaling Strategies

% When 32B Outperforms 670B: Rethinking Model Scaling Principles for Software Engineering Agents

% Unlocking Code Reasoning Potential in Deployable LLMs for Software Engineering Agents

% Thinking Longer, Not Larger: Test-Time Scaling for Efficient and Deployable Software Engineering Agents

% Thinking Longer, Not Larger: Enhancing Software Engineering Agents through Scaling LLM Test-Time Compute


\author{Yingwei Ma, Yongbin Li\textsuperscript{\dag}, Yihong Dong\textsuperscript{*}, Xue Jiang\textsuperscript{*}, Rongyu Cao, Jue Chen, Fei Huang, Binhua Li}
\thanks{\textsuperscript{\dag}Corresponding Author.}
\thanks{\textsuperscript{*}Work done during Yihong and Xue's internship at Tongyi Lab. Both are students at Peking University.}
\email{mayingwei.myw@alibaba-inc.com}
\affiliation{%
  \institution{Tongyi Lab, Alibaba Group}
  \city{Beijing}
  \country{China}
}

\renewcommand{\shortauthors}{Trovato et al.}

\begin{abstract}

Recent advancements in software engineering agents have demonstrated promising capabilities in automating program improvements. However, their reliance on closed-source or resource-intensive models introduces significant deployment challenges in private environments, prompting a critical question: \textit{How can personally deployable open-source LLMs (e.g., 32B models running on a single GPU) achieve comparable code reasoning performance?}

To this end, we propose a unified Test-Time Compute (TTC) scaling framework that leverages increased inference-time computation instead of larger models. Our framework incorporates two complementary strategies: internal TTC and external TTC. Internally, we introduce a \textit{development-contextualized trajectory synthesis} method leveraging real-world software repositories to bootstrap multi-stage reasoning processes, such as fault localization and patch generation. We further enhance trajectory quality through rejection sampling, rigorously evaluating trajectories along accuracy and complexity. Externally, we propose a novel \textit{development-process-based search} strategy guided by reward models and execution verification. This approach enables targeted computational allocation at critical development decision points, overcoming limitations of existing "end-point only" verification methods.

Evaluations on SWE-bench Verified demonstrate our \textbf{32B model achieves a 46\% issue resolution rate}, surpassing significantly larger models such as DeepSeek R1 671B and OpenAI o1. Additionally, we provide the empirical validation of the test-time scaling phenomenon within SWE agents, revealing that \textbf{models dynamically allocate more tokens to increasingly challenging problems}, effectively enhancing reasoning capabilities. We publicly release all training data, models, and code to facilitate future research.\footnote{\url{https://github.com/yingweima2022/SWE-Reasoner}}

\end{abstract}


% \ccsdesc[500]{Software and its engineering~Automatic programming}
% \ccsdesc[500]{Computing methodologies~Multi-agent planning}



\setcopyright{none} % to remove the copyright notice
\settopmatter{printacmref=false} % to remove the ACM Reference Format

\maketitle

\section{Introduction}
Large language model (LLM)-based agents have emerged as promising tools for automating various software engineering tasks, particularly in software maintenance (e.g., bug fixing) and evolution (e.g., adding new features). The SWE-bench~\citep{jimenez2023swe} has become a critical benchmark for evaluating the capabilities of SWE agents, specifically designed to simulate real-world software improvement tasks. Given a natural language description of an issue and the corresponding GitHub repository, SWE agent is tasked with generating a patch that resolves the issue. The typical framework in code agent research involves locating the relevant code, generating a patch, and verifying its correctness~\cite{autocoderover,yang2024sweagent,xia2024agentless}. 

The main driver of progress in the field has been scaling model parameters and training data, leading to notable improvements in model capabilities. However, this scaling introduces critical deployment challenges. For instance, DeepSeek V3 671B requires 436GB of VRAM, even with 4-bit quantization, and demands multi-GPU setups (e.g., 6 NVIDIA A100 80GB)~\citep{GPUrequirements}, making such systems impractical for most organizations. Additionally, closed-source models like Claude 3.5 raise privacy concerns when used via API services, particularly regarding private code repositories. These challenges lead to our central research question: \textit{How can we unlock the code reasoning potential of deployable LLMs, achieving comparable performance?} For example, the 4-bit quantized 32B model requires only 21GB of VRAM and can run on a single NVIDIA RTX4090 card~\citep{GPUrequirements}.

To address this challenge, we propose shifting the scaling paradigm from model size to increasing the inference time inspired by emerging Test-Time Compute Scaling approaches~\citep{o1systemcard, o3minisystemcard}. Current TTC implementations take two forms: Internal TTC, where models are trained to enhance reasoning depth through longer Chain-of-Thought (CoT); and External TTC, where multiple outputs are generated in parallel, and the optimal solution is selected using search-based strategies. Despite the potential of these approaches, technical difficulties, including resource constraints and proprietary strategies, have limited further exploration in this area. Specifically, the following issues remain underexplored:

\begin{itemize}
\item \textit{Proprietary Implementation Barriers}: While models like OpenAI o1~\citep{o1systemcard} and DeepSeek R1~\citep{guo2025deepseek} have demonstrated the effectiveness of long CoT reasoning, their methodologies remain proprietary and rely heavily on non-public training data and requires substantial computational resources and data collection efforts, making replication challenging. Given the privacy concern surrounding software repositories, there is a pressing need for transparent and computationally efficient methods, enabling strong reasoning capabilities even within resource-constrained, private development environments.
\item \textit{Search Strategies Limitations}: Existing external TTC approaches employ simplistic selection mechanisms like majority voting~\citep{xia2024agentless}, which prove inadequate for software tasks requiring precise understanding of development context. Few studies have systematically analyzed the impact of different search strategies—such as outcome and process reward models, or test-driven verification—on guiding the issue resolution process.
\end{itemize}

\textbf{Our Approach.} To answer these questions, we conduct a systematic exploration on the challenging SWE-bench Verified~\citep{swebenchverified} proposed by OpenAI. We build upon an open-source SWE framework (SWESyninfer~\citep{ma2024lingmaswegpt}) to generate initial single-solution proposals, which divide the issue resolution process into three key steps: (1) identifying relevant codebase context (repository understanding), (2) fault localization, and (3) generating candidate code edits. We then explore both internal and external TTC methods to enhance agent performance.

\textit{For internal TTC}, we propose a \textit{development-contextualized trajectory synthesis} method to address limited due to a lack of realistic multi-stage reasoning data aligned with actual software development workflows. Specifically, we first scrape <issue, repository, pull-request> triplets from high-quality GitHub repositories (>1000 stars) and construct executable verification environments; we then use DeepSeek R1 as a bootstrapping model to generate comprehensive reasoning trajectories spanning repository understanding, fault localization, patch generation, and patch verification. These trajectories are refined through \textit{Development-Contextualized Rejection Sampling}, which ensures quality via multi-dimensional filtering that evaluates both accuracy and complexity (filtering out problems solvable by small base model without refinement). Finally, our \textit{Reasoning Training} preserves both the think component (capturing planning, reflection, and correction processes) and the answer component (final solutions) at each reasoning step, enabling the model to internalize the multi-step decision-making process essential for complex software engineering tasks. This approach resolves 37.6\% of issues on SWE-bench Verified with trained 32B model, surpassing Llama 3.1 405B~\citep{llama3.1}. Our results demonstrate that smaller models can achieve comparable capabilities to much larger models when trained on high-quality, multi-step reasoning trajectories derived from real software development scenarios.

For \textit{External TTC}, we introduce a \textit{development-process-based search} strategy that strategically focuses computational resources on critical decision points in the software engineering workflow. Unlike existing approaches that either validate only at the final solution stage~\citep{pan2024swegym, xia2024agentless}, our framework applies targeted search at three crucial development phases: repository understanding, fault localization, and patch generation. We train specialized Process Reward Model (PRM) to evaluate intermediate outputs at these critical junctures, effectively pruning less promising solution paths early while maintaining a manageable beam width. At the patch generation stage, we implement execution verification through automatically generated reproduction code, providing concrete feedback on patch correctness. For final solution selection, we employ an Outcome Reward Model (ORM) trained via Direct Preference Optimization on verified patch pairs, enabling effective ranking of candidate solutions without requiring access to intermediate reasoning steps. Our experiments demonstrate that this development-process-based search strategy significantly improves performance with fixed model size, and when combined with our Internal TTC approach, yields even greater performance gains. These results highlight how strategic test-time computation allocation can achieve performance comparable to much larger models while maintaining computational efficiency. Additionally, we provide the first empirical validation of the test-time scaling phenomenon within SWE agents, revealing that models dynamically allocate more tokens to increasingly challenging problems, effectively enhancing reasoning capabilities.

\textbf{Contributions.} In summary, we make the following novel contributions:
\begin{itemize}
    \item We propose a unified scaling TTC approach tailored specifically for software engineering agents, including Internal TTC and External TTC.
    \item Our method achieves state-of-the-art open source results on the challenging SWE-bench Verified benchmark, resolving 46\% of issues with a 32B model. Notably, our approach surpasses larger models, demonstrating the effectiveness of targeted inference-time scaling.
    \item We present the empirical validation of the test-time scaling phenomenon within SWE agents, showing that increased inference-time computation improves performance on challenging software engineering problems.
    \item We open-source our model checkpoints, data, and code to support further research and development in this field.
\end{itemize}

\begin{figure*}
    \centering
    \includegraphics[scale=0.45]{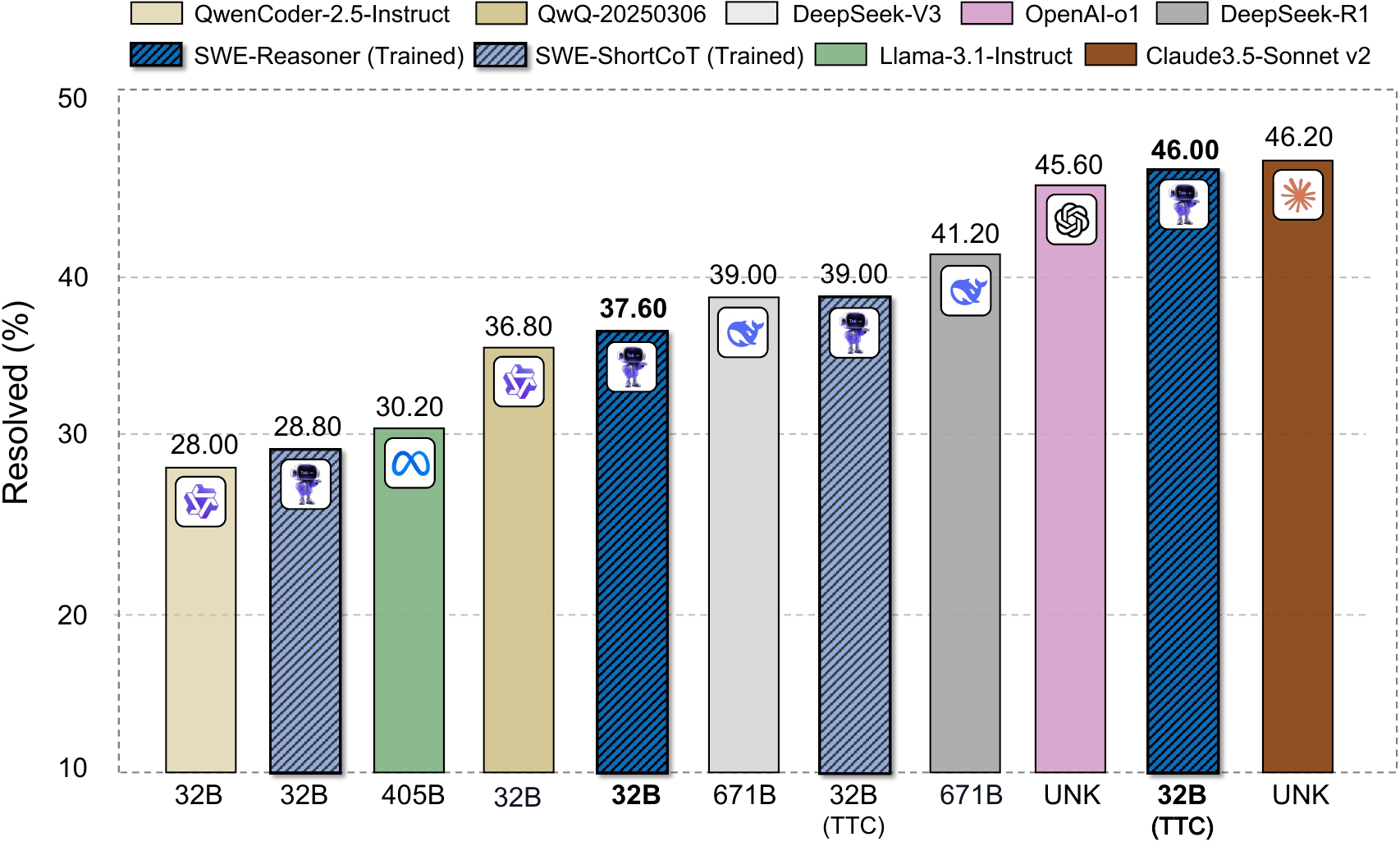}
    % \vspace{-1.0em}
    \caption{Comparison between the performance of smaller LLMs with extended Test-Time Compute and larger models on SWE-Bench Verified.}
    % \vspace{-1.0em}
    \label{fig:main_results}
\end{figure*}

\begin{figure*}
    \centering
    \includegraphics[scale=0.42]{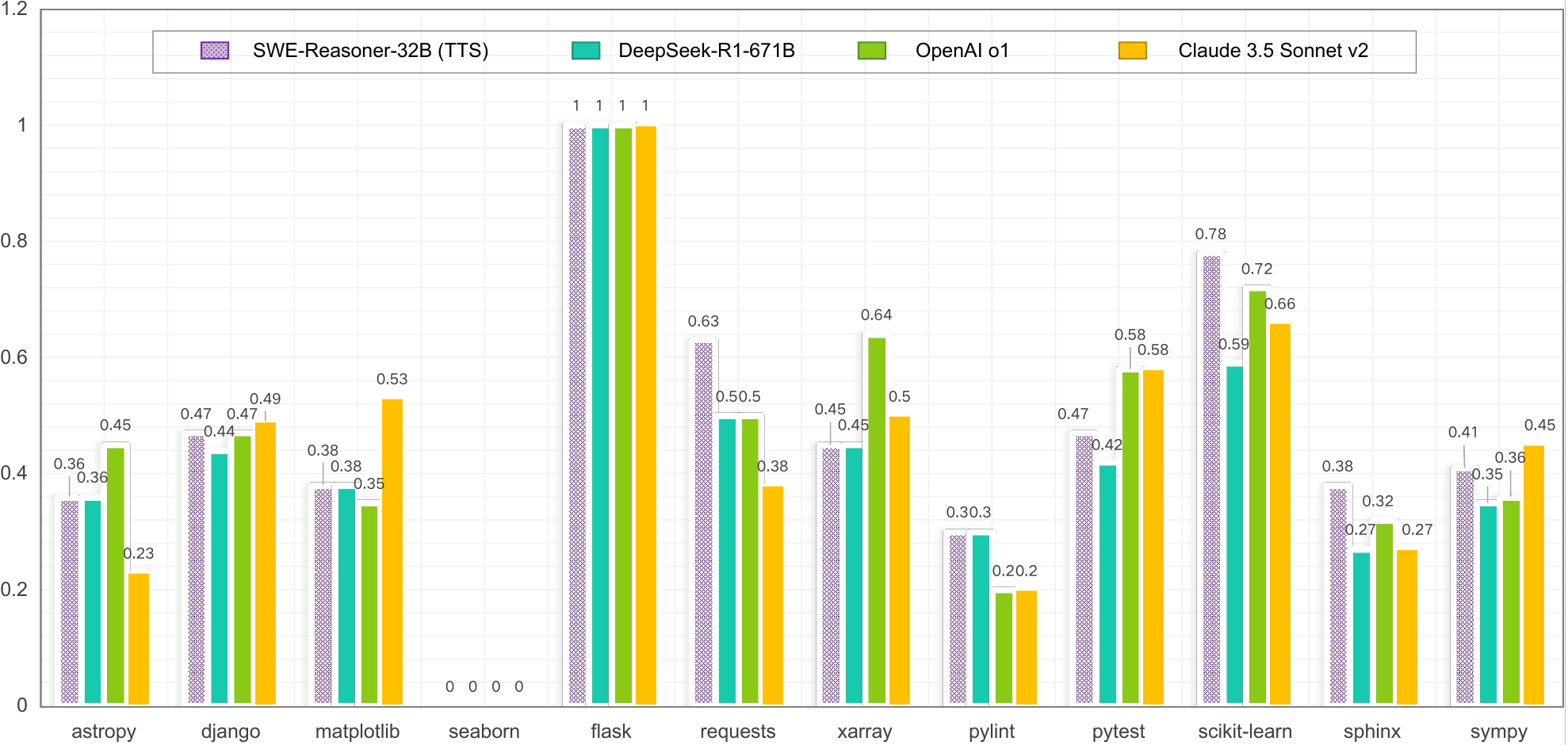}
    % \vspace{-1.0em}
    \caption{Comparison of issue resolution rates between our unified TTC framework (32B) and other LLMs across different repositories in SWE-bench Verified.}
    % \vspace{-1.0em}
    \label{fig:repo_performance}
\end{figure*}

\section{Test-Time Computation Explored: Internal and External Strategies}

\begin{figure*}
    \centering
    \includegraphics[scale=0.43]{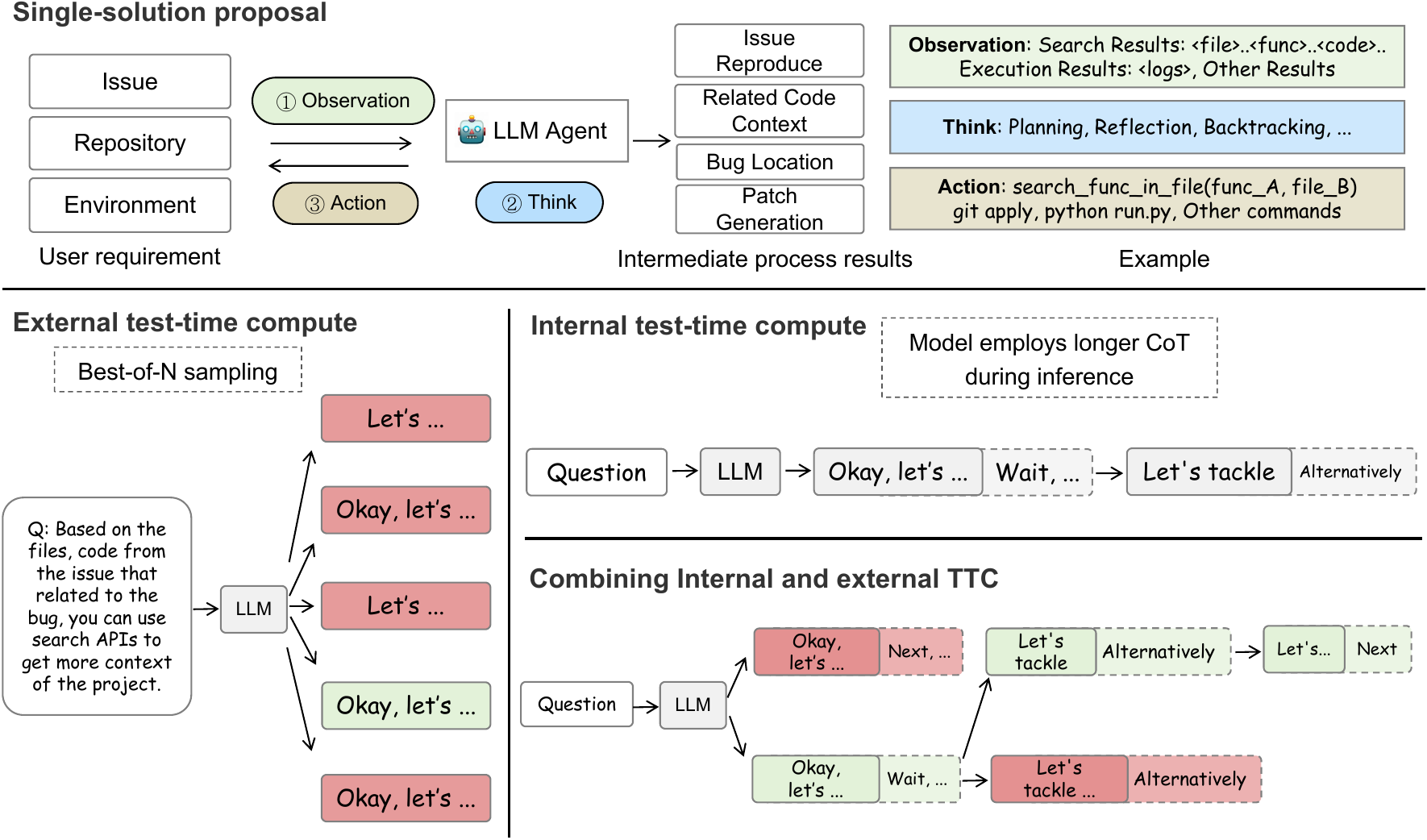}
    % \vspace{-1.0em}
    \caption{A unified view of Test-Time Scaling strategies for SWE agents. Internal TTC enhances reasoning depth through extended chain-of-thought training, while External TTC employs reward-guided search and verification to select optimal solutions. The hybrid approach combines both paradigms through iterative refinement.}
    % \vspace{-1.0em}
    \label{fig:overview}
\end{figure*}

In this section, we explore two core strategies for enhancing SWE-agent performance through scaling TTC: Internal and External TTC. Figure~\ref{fig:overview} presents our unified framework, illustrating how these approaches improve software engineering task. We first provide an overview of these two strategies and then delve into their specific implementations and results.

\subsection{Internal TTC in Software Engineering}

\textit{Internal TTC} aims to enhance the reasoning depth during inference by leveraging extended CoT. While OpenAI o1 and DeepSeek R1 achieve strong performance via large-scale Reinforcement Learning (RL) and massive datasets, we hypothesize that training smaller models (e.g., 32B parameters) using bootstrapped long reasoning trajectories, augmented by development-contextualized rejection sampling, can activate comparable reasoning capabilities. This is primarily because the model has already encoded a wealth of software engineering knowledge during pre-training. By utilizing high-quality, multi-step reasoning trajectories derived from real software development scenarios during post-training, we provide effective multi-step decision supervision, which helps unlock the model’s reasoning potential. To validate this, we introduce a systematic approach for synthesizing high-quality reasoning trajectories, consisting of three primary stages: data curation, trajectory bootstrapping, and development-contextualized rejection sampling.

\subsubsection{High-Quality Trajectory Synthesis} The foundation of our approach lies in high-quality, real-world software development data. In \textbf{Data Curation} stage, we begin by scraping <issue, pull-request, codebase> triplets from GitHub using SWE-bench's data collection procedure~\citep{jimenez2023swe}, focusing on repositories with high star ratings (>1000 stars) to ensure code quality. \textit{We filter out repositories already present in the SWE-bench dataset to avoid data leakage.} For each selected repository, we collect issues and linked pull requests (PRs) that were merged by developers. To further enhance the quality of the data, we apply a set of heuristic filtering rules, similar to those used in OctoPack~\cite{muennighoff2023octopack}. For issues, we retain only those with textual descriptions containing at least 20 characters to exclude overly vague or incomplete issues. Additionally, we filter out issues containing more than three hyperlinks, as these are often references to external resources rather than detailed descriptions of the issue at hand. For pull requests, we focus on those that modify between one and five code files, excluding those that only modify test files. This ensures that the changes are substantive. To ensure that each repository is suitable for patch verification, we use ExecutionAgent~\citep{executeagent} to automatically construct the execution environment, ensuring the necessary dependencies and execution contexts are properly set up. We filter out repositories where the environment cannot be built or run, resulting in a final dataset of 9,000 issues from 300 repositories, with verified executable environments capable of real-time patch validation.

In \textbf{Trajectory Bootstrapping} stage, we employ a bootstrapping strategy to synthesize detailed problem-solving trajectories. This approach builds upon the open-source SWE framework (SWE-SynInfer~\citep{ma2024lingmaswegpt}), which has achieved superior results in open-source models. SWE-SynInfer divides the issue resolution process into three steps: (1) repository understanding to identify relevant codebase files, (2) fault localization to pinpoint problematic code segments, and (3) patch generation to produce candidate code edits. We extend this framework to include a Patch Verification phase, following Agentless~\citep{xia2024agentless}, and call it \textit{SWE-SynInfer+}. In this enhanced phase, the model generates reproduction code based on the issue description, and then verifies the correctness of the generated patch by executing the reproduction code. If the patch is deemed incorrect, the model iterates, refining the solution until it either meets the verification criteria or reaches the maximum threshold of iterations. \label{reproduce_code_chapter}
We use a open-source reasoning model (DeepSeek R1~\citep{guo2025deepseek}) to bootstrap these long reasoning trajectories, as R1 iterates and refines its internal reasoning multiple times by utilizing more inference computation before producing the final output. Each trajectory step in the bootstrapping process includes two primary components: the \textit{think} component, which captures the planning, reflection, and correction processes, and the \textit{answer} component, which represents the final solution for that step. The trajectory bootstrapping process is summarized in Algorithm \ref{alg:bootstrapping_revised}, which outlines how the model generates a sequence of reasoning steps. This algorithm mirrors real-world software development practices, where each stage builds upon previous reasoning in an iterative manner, progressively refining the solution. The environment is updated as the reasoning process progresses, and the model continues until either the patch is successfully verified by reproduce code or the maximum number of steps is reached.

\begin{algorithm}[t]
\caption{Trajectory Bootstrapping Process}
\label{alg:bootstrapping_revised}
\begin{algorithmic}[1]
\State \textbf{Input}: Issue $I$, Repository $R$, Base Model $M$
\State Initialize trajectory $\tau = []$, Environment $\mathcal{E}$
\Procedure{GenerateTrajectory}{$I$, $R$, $M$}
\For{step $t \in \{1, \dots, T_{\max}\}$}
    \State $s^{t}_{\text{think}}, s^{t}_{\text{answer}} \leftarrow M(\text{CoT-Prompt}(I, R, \tau[1 : t-1]))$ 
    \State \texttt{ActionType, Params} $\leftarrow$ \Call{Analyze}{$s^{t}_{\text{answer}}$}
    \If{parsing failed}
        \State $\tau$.append(fallback\_error\_handling)
        \State \textbf{continue}
    \EndIf
    \State $s^{t}_{\text{output}} \leftarrow$ \Call{ExecuteAction}{\texttt{ActionType, Params}, $\mathcal{E}$}
    \State $\tau$.append$\left( \left( (s^{t}_{\text{think}}, s^{t}_{\text{answer}}), s^{t}_{\text{output}} \right) \right)$
    \State Update $\mathcal{E}$ with $s^{t}_{\text{output}}$ outcomes
    \If{$\text{Resolved}(\mathcal{E})$ \textbf{or} $\text{Failed}(\mathcal{E})$}
        \State \textbf{break}
    \EndIf
\EndFor
\State \Return $\tau$
\EndProcedure
\end{algorithmic}
\end{algorithm}

We use \textbf{Development-Contextualized Rejection Sampling} to ensuring the quality of generated reasoning trajectories, which contain \textbf{accuracy} and \textbf{complexity} of each trajectory.
\begin{itemize}
    \item Repository Understanding: We verify that the model correctly identifies the files that need modification. Specifically, we compare the model’s output in the Repository Understanding phase with the files changed in the developer’s patch, ensuring alignment with the actual code modifications.

    \item Fault Localization: The generated patch must focus on the correct locations within the code (e.g., relevant classes, functions, and surrounding code blocks). We check that the model’s patch includes changes at these same locations as those in the developer’s patch.
    
    \item Issue Reproduce: We validate the generated reproduction code's correctness against the developer's patch. A valid reproduction code should output \textit{issue reproduced} when executed on the original codebase and \textit{issue resolved} when executed after applying the developer's patch. This two-stage verification ensures that the reproduction code correctly captures the essence of the issue and can reliably detect when the issue has been fixed.

    \item Patch Correctness: We assess whether the patch resolves the issue. We apply the model’s patch to the repository and run the SWE agent’s reproduction code to check if the issue is fixed. For cases where the LLM fails to generate correct reproduction code, we follow the approach~\cite{ma2024lingmaswegpt} by evaluating the similarity between the model-generated patch and the developer's patch as a filtering criterion. We also run existing unit tests to ensure the patch does not break other functionalities, verifying the correctness and stability of the solution.

    \item Complexity Filtering: To focus on challenging problems that activate deeper reasoning capabilities, we filter out simpler issues that Qwen2.5 Coder 32B~\citep{hui2024qwen2.5-coder} can solve in a single attempt without refinement. This ensures our training data consists of problems requiring sophisticated long CoT reasoning.

\end{itemize}

By incorporating development context into the rejection sampling process, we ensure that only high-quality trajectories are retained, ultimately enhancing the model's reasoning depth and performance. Additionally, if a patch is incorrect but the preceding reasoning stages are accurate, we discard the erroneous patch data while preserving the correct stage data. This allows us to retain valuable reasoning steps, ensuring that useful problem-solving knowledge is not lost during the filtering process.

\subsubsection{Training} We train our model using supervised learning on the synthesized long CoT trajectories dataset. Our objective is to enable the model to internalize structured multi-round reasoning. We follow a standard maximum likelihood estimation objective, optimizing the conditional probability of generating correct reasoning actions given an issue and prior observations. The training loss is computed over both the \textit{think} and \textit{answer} components at each step, ensuring that the model learns both intermediate reasoning steps and final predictions. To enhance efficiency in multi-round inference, we adopt a history pruning mechanism inspired by DeepSeek R1~\citep{guo2025deepseek}. Specifically, for each reasoning step $i$, we discard the \textit{think} component of the previous response and retain only the final \textit{answer} in the historical context. Formally, given a training instance consisting of issue and the corresponding step-wise trajectory: 

\begin{equation}
\theta' \leftarrow \argmax_\theta \sum_{(s_{\text{obs}}^i, s_{\text{think}}^i, s_{\text{ans}}^i) \in \text{traj}} \log P_\theta(s_{\text{think}}^i, s_{\text{ans}}^i \mid \text{issue}, s_{\text{obs}}^i, \mathcal{H}_{i-1})
\end{equation}

\begin{equation}
\mathcal{H}_i = \mathcal{H}_{i-1} \cup \{ s_{\text{obs}}^i, s_{\text{ans}}^{i-1} \}
\end{equation}

where \( s_{\text{obs}}^i \) represents the structured observations at step \( i \), capturing relevant code snippets, execution logs, or other extracted information crucial for reasoning. \( s_{\text{think}}^i \) represents the model’s internal reasoning process, and \( s_{\text{ans}}^i \) represents the actionable output from the model at each step, such as the search\_api, the specific patch to apply, or a command to run.
\( \mathcal{H}_{i-1} \) denotes the historical trajectory context up to step \( i-1 \), ensuring that the model conditions on prior reasoning states when generating the next step.

\subsection{Effective Search Strategies for External TTC}

\textit{External TTC} explores ways to leverage multiple inference outputs to identify the best solution (see Figure~\ref{fig:overview}). Existing methods typically generate several candidate patches at once and then rely solely on a final correctness check (e.g., by running unit tests~\citep{yang2024sweagent, swerl}, regression tests~\citep{xia2024agentless}, or outcome-based reward models~\citep{pan2024swegym}) to select the best candidate. However, such “end-point only” methods often underutilize the available search budget because they do not intervene intermediate reasoning steps. This is particularly problematic for SE tasks, which involve lengthy reasoning chains with multiple interdependent decisions. Moreover, classical tree search~\citep{nebius} (like beam search) applied at \emph{every} intermediate step (i.e., “step-by-step” validation) is also infeasible for extensive software development pipelines, due to the computational overhead of verifying. To address this, we propose a development-process-based search strategy that focuses on the critical decision phases of software development.

\subsubsection{Development-Process-Based Search Strategy} 
\begin{figure}
    \centering
    \includegraphics[scale=0.35]{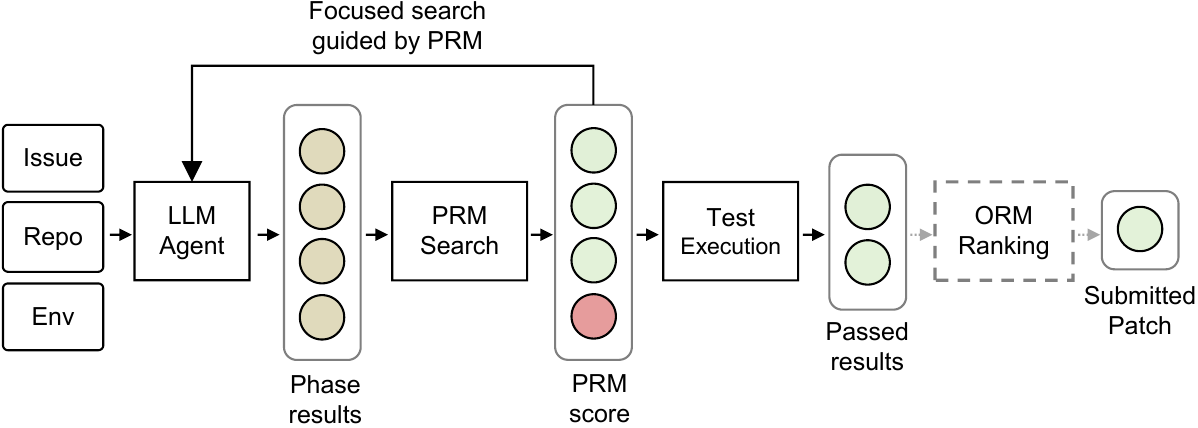}
    % \vspace{-1.0em}
    \caption{Overview of Development-Process-Based Search Strategy.}
    % \vspace{-1.0em}
    \label{fig:dev_search}
\end{figure}

We decompose the agent's problem-solving process into three essential phases: (1) repository understanding, (2) fault localization, and (3) patch generation. These phases represent crucial decision points in the development process, where errors can propagate and dramatically affect subsequent steps. By focusing our search at these junctures, we ensure that the agent's decisions are evaluated at critical stages, not at every single action within the process. Figure \ref{fig:dev_search} presents our overview framework.

\textbf{Focused Search with Process Reward Model (PRM)}. At the repository understanding and fault localization stages, we apply a lightweight beam search strategy, guided by PRM. For each stage, we generate N candidate outputs and use the PRM to score each candidate based on its likelihood of correctness. The top-k highest-scoring candidates are retained and used as input for the next stage, effectively pruning less promising solution paths. This approach maintains a manageable beam width while focusing computational resources on the most promising solution trajectories. 

\textbf{Patch Generation and Execution Verification}. At this stage, the agent generates potential patches to resolve the identified bug. To ensure the correctness of the generated patches, we apply execution verification, where the agent generates reproduction code to check if the patch successfully fixes the issue. This verification process also ensures that the patch does not introduce new bugs by running regression tests on the repository to confirm that existing functionality is unaffected.

\textbf{Final Ranking with Outcome Reward Model (ORM)}. After executing the verification checks, we prioritize keeping patches that pass more tests, and then we select the most promising patches from among these (in case of a tie). Here, we apply the ORM, which evaluates the quality of the final patches. The ORM ranks multiple candidate patches and the highest score from the ORM is selected as the final solution to be submitted. Importantly, our ORM design requires only the issue description and the candidate patch as inputs, without depending on intermediate reasoning steps or specific agent architectures. This design choice ensures that our ORM can be seamlessly integrated with various SWE agent systems or CI/CD pipelines.

\subsubsection{Reward Model Training}

\textbf{Process Reward Model (PRM)} The PRM aims to assess the intermediate correctness at critical development phases, namely repository understanding and fault localization. To train the PRM, we construct a labeled dataset by leveraging the high-quality bootstrapped trajectories generated during the trajectory synthesis phase. For each trajectory step, we formulate a binary classification task where the PRM learns to distinguish between correct and incorrect intermediate outputs. Specifically, for repository understanding, the model predicts whether the identified files align with the actual developer’s modified files. For fault localization, it predicts whether the model-generated patch aligns with the developer-edited detailed locations. We use the contextual information from issues and intermediate trajectory reasoning outputs as inputs, enabling the PRM to contextualize and effectively evaluate partial solutions. We fine-tune a base model using a standard next-token prediction objective with cross-entropy loss, guiding the model to output tokens corresponding to binary labels (i.e., “+” for correct and “-” for incorrect):

\begin{equation}
\mathcal{L}_{PRM} = -\sum_{i=1}^{N} [y_i \log(p_i) + (1 - y_i) \log(1 - p_i)]
\end{equation}

where $y_i$ is the binary correctness label (1 for correct, 0 for incorrect), and $p_i$ is the PRM's predicted probability of correctness.

\textbf{Outcome Reward Model (ORM).} The ORM performs final sorting of the generated patches. For ORM training, we curate a dataset comprising pairs of candidate patches labeled according to their verification outcomes. Specifically, patches that pass all execution verification and regression tests are considered superior (winning response), while those failing any verification steps are inferior (losing response). To effectively capture relative patch quality, we apply the Direct Preference Optimization (DPO) loss~\citep{rafailov2023direct} for training:

\begin{equation}
\begin{aligned}
\mathcal{L}_{ORM}(\pi_\theta; \pi_{\text{ref}}) = -\mathbb{E}_{(x,y_w,y_l)} \Bigg[ \log \sigma \Bigg(\beta \log \frac{\pi_\theta(y_w \mid x)}{\pi_{\text{ref}}(y_w \mid x)} - \\
\beta \log \frac{\pi_\theta(y_l \mid x)}{\pi_{\text{ref}}(y_l \mid x)}\Bigg) \Bigg]
\end{aligned}
\end{equation}

Here, $y_w$ represents the winning patch (passes verification), $y_l$ is the losing patch (fails verification), and $x$ is the associated issue description. We fine-tune a smaller base model as the ORM reference model ($\pi_{\text{ref}}$) to maintain fast inference during the external TTC phase. The hyperparameter $\beta$ controls the reward sharpness, and we chose a common value of 0.5.

\subsection{Putting It Together} We propose a unified framework by seamlessly integrating internal and external Test-Time Scaling (TTC), emphasizing enhanced performance of software engineering agents through allowing models to \textit{think longer} and \textit{search more}, instead of increasing model size. Figure \ref{fig:overview} illustrates this unified TTC framework, clearly demonstrating the integration of internal and external scaling strategies. All models in our experiments are based on Qwen2.5 Coder 32B~\citep{hui2024qwen2.5-coder}. Our approach ultimately shows that careful inference-time scaling can achieve or even surpass the performance of significantly larger models, thus enabling advanced software engineering reasoning capabilities even under constrained computational resources. The effectiveness of this approach will be thoroughly validated through subsequent experiments.

\section{Evaluation}

\subsection{Benchmark and Evaluation Metric}

\textit{SWE-bench Verified.} We evaluated our method on the recently proposed benchmarks SWE-bench Verified~\cite{swebenchverified}, comprising 500 real-world GitHub issues. The model receives only the natural language description of the original GitHub issue and its corresponding code repository as input. These benchmarks employ developer-written unit tests to verify the correctness of model-generated patches, ensuring a rigorous assessment of the model's performance. 

\textit{Evaluation Metric.} We use (1) the percentage of resolved task instances, (2) fault location success rate. These evaluation metrics represent overall effectiveness in resolving real-world GitHub issues. In addition, we evaluate the effectiveness of solving issues at different difficulty levels and different generation budgets to verify the test-time scaling phenomenon of our method.

\subsection{Overall Effectiveness of Unified TTC Framework}

\begin{table}
\renewcommand{\arraystretch}{1.0}
\centering
\scalebox{1.0}{
\begin{tabular}{llc}
\hline
\textbf{Agent} & \textbf{LLM} & \textbf{Verified} \\
\hline
\rowcolor{gray!25}
\multicolumn{3}{c}{\includegraphics[height=1em]{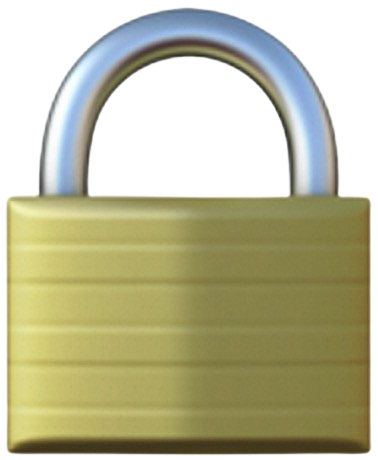} Model size unknown or size $>$ 100B} \\ \hline 
SWE-agent~\cite{swebenchverified} &  GPT-4o & 23.00\%  \\

AutoCodeRover~\cite{swebenchverified} &  GPT-4o & 28.80\% \\ 
SWE-SynInfer~\citep{ma2024lingmaswegpt} &  GPT-4o & 31.80\%  \\	
Agentless~\cite{swebenchverified} &  GPT-4o & 33.20\%\\
SWE-agent~\cite{yang2024sweagent} &  Claude3.5-Sonnet-v1 & 33.60\%  \\ 	
SWE-SynInfer~\citep{ma2024lingmaswegpt} & Claude3.5-Sonnet-v1 & 35.40\%  \\
OpenAI Tools~\citep{gpt4_5systemcard} & GPT-4.5 & 38.00\% \\

Agentless~\citep{o3minisystemcard} & OpenAI-o3-mini & 40.00\% \\
Agentless~\citep{o1systemcard} &  OpenAI-o1-1217 & 41.00\% \\ 

% SWE-SynInfer+ & DeepSeek-R1 & 41.20\%  \\

% SWE-SynInfer+ & Claude3.5-Sonnet-v2 & 46.20\%  \\
Anthropic Tools~\citep{claude3.5} & Claude3.5-Sonnet-v2 & 49.00\%  \\
OpenAI Tools~\citep{o3minisystemcard} & OpenAI-o3-mini & 61.00\% \\
\textbf{Anthropic Tools}~\citep{clade3.7} & \textbf{Claude3.7-Sonnet} & \textbf{62.30\%} \\
\hline

\rowcolor{gray!25}
\multicolumn{3}{c}{\includegraphics[height=1em]{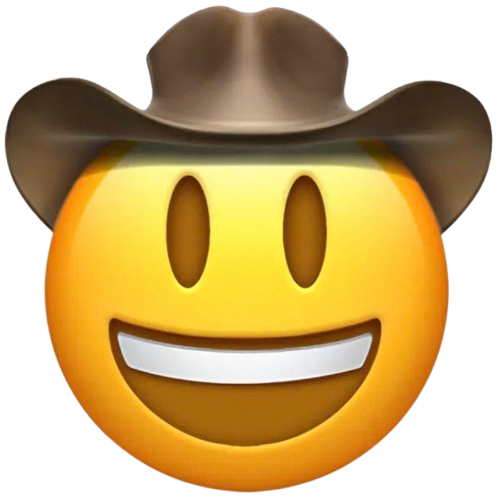} Model size $\leq$ 100B} \\ \hline 
Agentless~\citep{ma2025sorft} & Qwen2.5-Coder 32B & 25.60\% \\ 
% SWE-SynInfer+ & Qwen2.5-Coder 32B & 28.00\% \\
SWE-Gym~\citep{pan2024swegym} &  SWE-Gym 32B & 29.80\% \\
SWE-SynInfer~\citep{ma2024lingmaswegpt} &  SWE-GPT 72B & 30.20\% \\ 
Agentless~\citep{ma2025sorft} &  SoRFT-Qwen 32B & 30.80\% \\ 
SWE-Fixer~\citep{ma2024lingmaswegpt} &  SWE-Fixer 72B & 32.80\% \\ 
NebiusAI~\citep{nebius} & NebiusAI 72B\&70B & 40.60\% \\
Agentless Mini~\citep{swerl} &  Llama3-SWE-RL 70B & 41.00\% \\
\textbf{SWE-SynInfer+} &  \textbf{SWE-Reasoner 32B} & \textbf{46.00\%} \\
\hline
\end{tabular}}
\caption{Performance comparison of our method and other models on SWE-bench Verified benchmark.}
\label{tab:main_results}
\end{table}

% \begin{table}[htbp]
%     \centering
%     \caption{SWE-Bench (OpenHands), Evaluated by xingyao}
%     \begin{tabular}{lccc}
%     \hline
%     \textbf{Model} & \textbf{mode} & \textbf{Lite} & \textbf{Verified} \\
%     \hline
%     \textbf{claude-3-7-sonnet} & function & - & 54.80\% \\
%     claude-3-5-sonnet-20241022 & function & 41.67\% & 53.00\% \\
%     claude-3-5-haiku-20241022 & function & 28.67\% & - \\
%     o3-mini-2025-01-31 & function & - & 43.70\% \\
%     o1-2024-12-17 & function & 22.00\% & 28.80\% \\
%     gpt-4o-2024-05-13 & function & 19.00\% & - \\
%     gpt-4o-2024-08-06 & function & 18.67\% & - \\
%     gpt-4o-mini-2024-07-18 & function & 5.67\% & - \\
%     Llama 3.1 70B (native FC) & function & 3.67\% & - \\
%     gemini-1.5-pro-002 & function & 1.00\% & - \\ \hline
%     Deepseek V3 & prompt & 23.00\% & 32.40\% \\
%     Deepseek-R1 & prompt & - & 34.00\% \\
%     gemini-1.5-pro-002 & prompt & 14.00\% & - \\
%     gpt-4o-2024-05-13 & prompt & 12.00\% & 25.00\% \\
%     Llama3.1 405B & prompt & 11.33\% & - \\
%     Llama3.3 70B & prompt & 10.67\% & - \\
%     Llama3.1 70B & prompt & 8.00\% & - \\
%     \textbf{Qwen2.5-Max} & prompt & - & 20.20\% \\
%     \textbf{Qwen2.5-72B-Instruct} & prompt & 7.00\% & - \\
%     \textbf{Qwen2.5-Coder-32B-Instruct} & prompt & 3.33\% & 7.00\% \\
%     gpt-4o-mini-2024-07-18 & prompt & 2.67\% & - \\

%     \hline
%     \end{tabular}
%     \label{tab:swe_bench_results}
% \end{table}

We evaluate the effectiveness of our unified TTC framework on the SWE-bench Verified benchmark. We first assess various base models under our SWE-SynInfer+ framework. Figure~\ref{fig:main_results} illustrates the comparative performance results. Notably, our 32B SWE-Reasoner model, which employs Internal TTC strategies, achieves an issue-resolution accuracy of 37.60\%. When combined with External TTC (budget=8), our model's performance further increases to 46.00\%. This unified approach closely matches the performance of the significantly larger proprietary Claude 3.5 Sonnet v2 model (46.20\%) and surpasses OpenAI-o1 (45.60\%) and DeepSeek-R1 (41.20\%), clearly demonstrating the effectiveness of our unfied TTC strategies. We further benchmark our approach against leading state-of-the-art SWE agent frameworks reported in existing literature (see Table~\ref{tab:main_results}). Within the $\leq$ 100B model-size category, our method achieves the highest issue-resolution accuracy, establishing a new state-of-the-art. Importantly, our method achieves this performance with substantially lower computational demands, emphasizing that careful inference-time computation strategies effectively leverage smaller models to reach competitive results.

Additionally, to evaluate the generalization and robustness of our unified TTC framework across different software domains, we analyzed its performance on a diverse set of repositories. Figure~\ref{fig:repo_performance} illustrates the issue-resolution rates of our SWE-Reasoner-32B (TTC) model across 12 representative software repositories, compared with the strongest open-source baseline (DeepSeek-R1 671B) and two leading closed-source models (OpenAI-o1 and Claude 3.5 Sonnet v2). Notably, SWE-Reasoner-32B (TTC) matches or surpasses the performance of DeepSeek-R1 671B in the majority of repositories, and closely approaches the performance of larger closed-source models in numerous instances. This consistent cross-domain performance underscores our method's robust generalization capabilities, highlighting its potential applicability and effectiveness across a wide spectrum of real-world software engineering scenarios.

\begin{figure}
    \centering
    \includegraphics[scale=0.63]{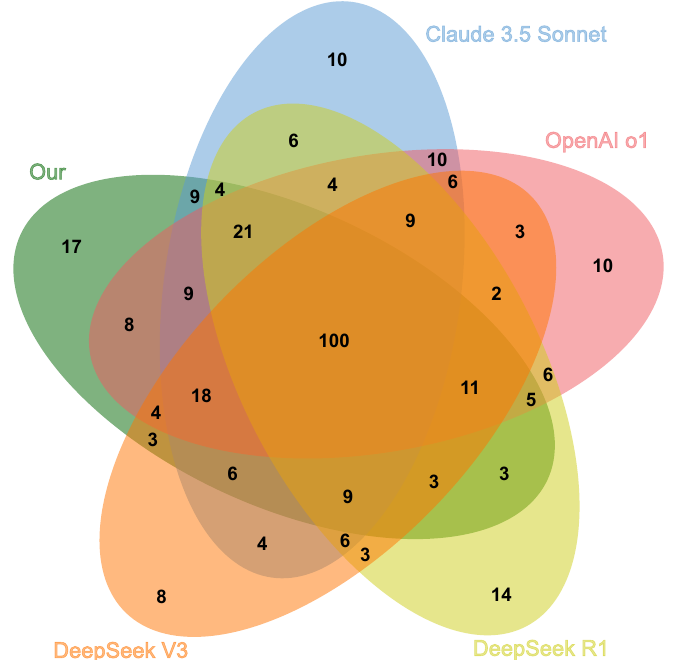}
    % \vspace{-1.0em}
    \caption{Venn diagram of issue instances solved by our unified TTC framework and other models on SWE-bench Verified.}
    % \vspace{-1.0em}
    \label{fig:venn}
\end{figure}

As illustrated in Figure~\ref{fig:venn}, we further analyzed the overlap of solved issue instances among different models on the SWE-bench Verified benchmark through a Venn diagram. The diagram reveals that our unified TTC framework uniquely solves 17 issue instances that other models fail to address, while also sharing a substantial number of successfully resolved issues with major models. This indicates that our approach not only achieves competitive performance quantitatively but also demonstrates unique problem-solving capabilities in terms of coverage.

\subsection{Analysis of Internal TTC Strategies}

We conducted two detailed analyses to comprehensively assess the effectiveness of our Internal TTC strategies:

\textbf{Effectiveness of Internal TTC via Ablation Study.} We performed ablation studies to assess the individual contributions of key Internal TTC components, particularly evaluating their impact on issue-resolution rates and fault localization accuracy. The results are summarized in Table \ref{tab:Ablation_InternalTTC}. Upon removing the Long Chain-of-Thought (Long CoT) component (\textit{-w/o. LongCoT}), we observed a significant reduction in issue resolution accuracy from 37.60\% to 28.80\%. Specifically, in the \textit{-w/o. LongCoT} experiment, we omitted the \textit{think} labels from training data, instead prompting Claude 3.5 Sonnet v2~\citep{claude3.5} to explicitly generate short-CoT reasoning and corresponding action predictions on the same dataset. We then applied the repository-aware rejection sampling method to this short-CoT data and trained the same base model (Qwen2.5-Coder 32B~\citep{hui2024qwen2.5-coder}). Despite leveraging the stronger Claude model for short-CoT generation, the trained smaller model underperformed compared to our original Long CoT strategy. This result highlights the unique advantage of Long CoT in activating deeper reasoning capabilities in smaller models. Additionally, we evaluated the impact of our repository-aware rejection sampling method by removing this filtering step (\textit{-w/o. Rejection}). Although using unfiltered synthesized data increased the overall volume of training data, issue-resolution performance decreased from 37.60\% to 33.00\%. This decline underscores the importance of carefully curated, high-quality reasoning trajectories for effective training.

\begin{table}
\renewcommand{\arraystretch}{1.2}
\centering
\scalebox{1.0}{
\begin{tabular}{lcccc}
\hline
\textbf{Ablation} & \textbf{Resolved} & \textbf{Chunk} & \textbf{Func} & \textbf{File} \\
\hline
\textbf{SWE-Reasoner} & \textbf{37.60\%} & \textbf{51.00\%} & \textbf{54.49\%} & \textbf{72.19\%}  \\
 -w/o. LongCoT & 28.80\% & 49.05\% & 51.68\% & 69.18\% \\ 
 -w/o. Rejection & 33.00\% & 48.76\% & 51.94\% & 71.38\%  \\  
 -w/o. All & 28.00\% & 44.22\% & 47.25\% & 60.69\% \\ \hline

\end{tabular}}
\caption{Ablation experiment of the Internal TTC method, where Resolved is the issue resolution rate on SWE-bench Verified, and Chunk, Func, and File are the fault location success rates at three different levels.}
\label{tab:Ablation_InternalTTC}
\end{table}

\begin{figure}
    \centering
    \includegraphics[scale=0.38]{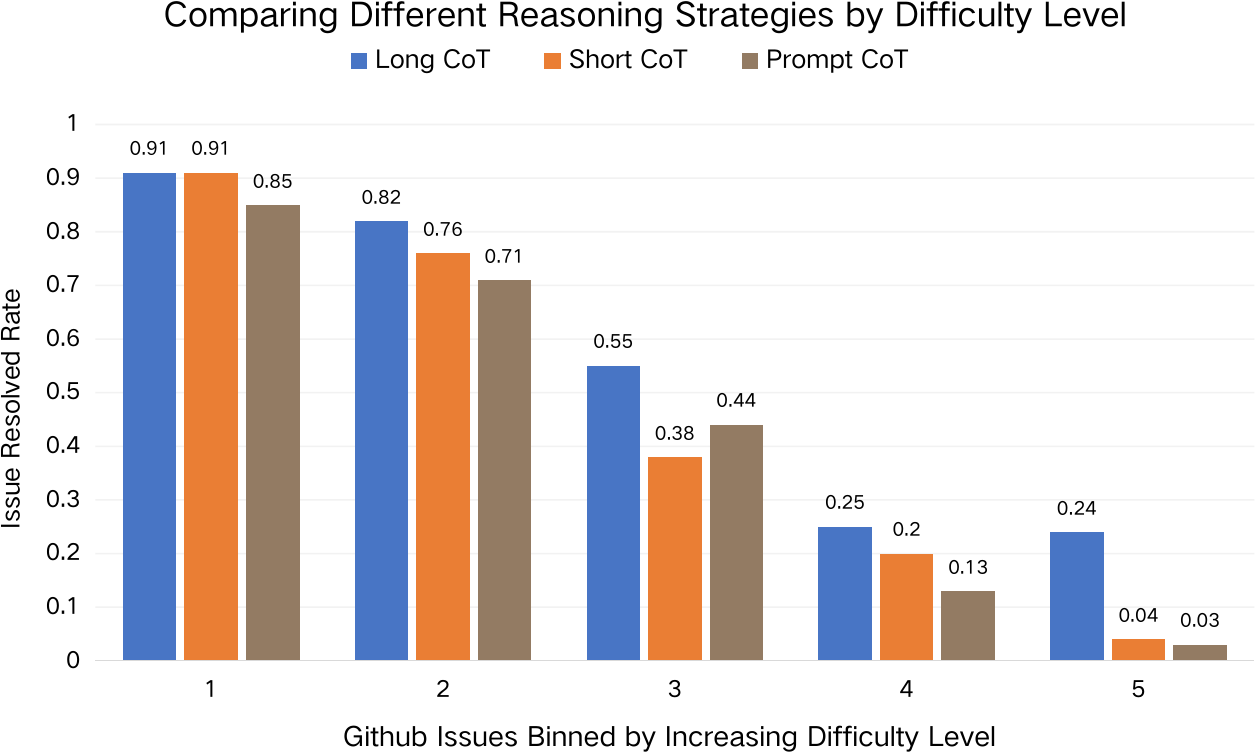}
    % \vspace{-1.0em}
    \caption{Comparison of Issue Resolution Rates by Reasoning Strategies across Difficulty Levels. The graph shows the performance of three approaches: Long CoT (SWE-Reasoner), Short CoT (\textit{w/o. Long CoT}), and Prompt CoT (\textit{w/o. All}).}
    % \vspace{-1.0em}
    \label{fig:cotsource}
\end{figure}

To further clarify the benefits of explicitly Long CoT reasoning trajectories training, we compared performance across internalized Long CoT (SWE-Reasoner), internalized Short CoT (\textit{w/o. Long CoT}), and prompt-based CoT (\textit{w/o. All}). Specifically, we categorized SWE-bench Verified issues into five difficulty buckets based on their resolution frequency among the top 30 submissions on the SWE-bench leaderboard~\citep{jimenez2023swe}. Level 1 includes issues resolved by 25–30 agent submissions (easiest), level 2 by 20–25 submissions, level 3 by 15–20 submissions, level 4 by 10–15 submissions, and level 5 by 5–10 submissions (hardest). Issues resolved fewer than five times were excluded due to their infrequency and high variance. As shown in Figure~\ref{fig:cotsource}, models employing internalized CoT (both Long and Short) consistently outperform Prompt-CoT-based methods. Crucially, our internalized Long CoT approach significantly surpasses Short CoT performance on the hardest bucket (level 5), achieving an issue-resolution rate approximately six times higher. These findings confirm that explicitly internalizing long reasoning trajectories is highly effective, particularly in enabling small models to tackle complex tasks by effectively leveraging test-time computational resources.

\begin{figure}
    \centering
    \includegraphics[scale=0.34]{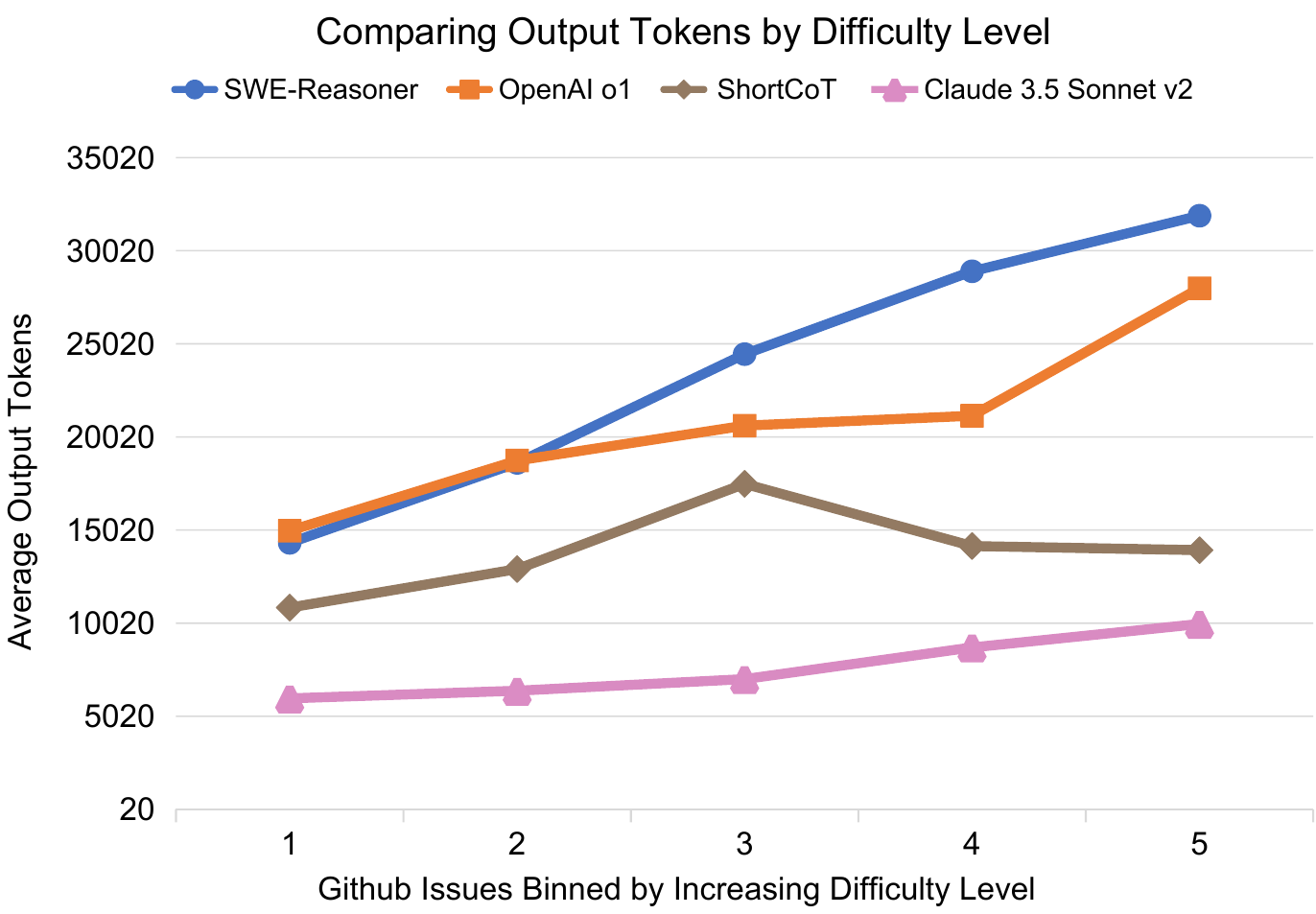}
    % \vspace{-1.0em}
    \caption{Average Number of Output Tokens by Difficulty Level. We categorize SWE-bench Verified issues into five difficulty buckets based on their resolution frequency by top-performing agents (bucket 1: resolved by 25–30 agents, bucket 5: resolved by 5–10 agents).}
    % \vspace{-1.0em}
    \label{fig:difficlty}
\end{figure}

\textbf{Analysis of the Test-Time Scaling Phenomenon.} We further investigated whether the SWE-Reasoner dynamically allocates computational resources based on task complexity, as indicated by longer inference trajectories (measured by output token counts). Using the previously defined difficulty buckets (level 1 being easiest and level 5 hardest), we compared average output tokens generated by SWE-Reasoner, OpenAI o1, ShortCoT model, and Claude 3.5 Sonnet v2 across different issue-difficulty levels (Figure~\ref{fig:difficlty}). From the figure, we observe that both SWE-Reasoner and OpenAI o1 continue to adaptively allocate more reasoning tokens to increasingly challenging tasks, demonstrating a clear test-time computation scaling phenomenon. Interestingly, we also observed that Claude3.5 Sonnet v2, despite being a model not explicitly trained for inference-intensive computation, exhibited a similar scaling trend, whereas ShortCoT model did not show this behavior clearly. This empirical evidence strongly supports the existence of test-time compute scaling in advanced reasoning models, further validating that our Internal TTC strategy effectively enables dynamic computational resource allocation tailored to task complexity.

\subsection{Analysis of External TTC Strategies}

We further evaluated the effectiveness of our proposed External Test-Time Compute (TTC) strategies, specifically the Development-Process-Based Search Strategy, through two targeted experiments. Due to the computational resources and significant time required for scaling experiments, we randomly sampled 100 issues from the SWE-bench Verified benchmark for these analyses.

\textbf{Effectiveness of Development-Process-Based Search Strategy.}
Our first experiment aimed to systematically evaluate the effectiveness of our proposed external search strategy, labeled as \textbf{Dev-Search}, against three alternative baselines under varying inference budgets (Generation Budget). Specifically, Dev-Search utilizes our proposed Process Reward Model (PRM)-guided beam search at the repository understanding and fault localization stages, combined with execution-based patch verification and ORM-based final ranking. For budgets of 2 and 4, we set the beam search width to 2; for budget 8, we expanded it to 4. We compared Dev-Search with the following baselines. \textbf{Exec} strategy uses only execution verification (regression tests and issue reproduction) to select a final patch. If multiple patches passed execution verification, a patch was randomly selected to resolve the tie.
In \textbf{ORM\_Exec} approach, we employed our Outcome Reward Model (ORM) to break ties among multiple patches that passed execution verification, rather than selecting randomly.
In \textbf{Voting} strategy, following Agentless~\cite{xia2024agentless}, we normalized patches to abstract syntax tree representations, standardized their format (ignoring comments, extra whitespace, and surface-level differences), and then selected the patch appearing most frequently.

\begin{figure}
    \centering
    \includegraphics[scale=0.36]{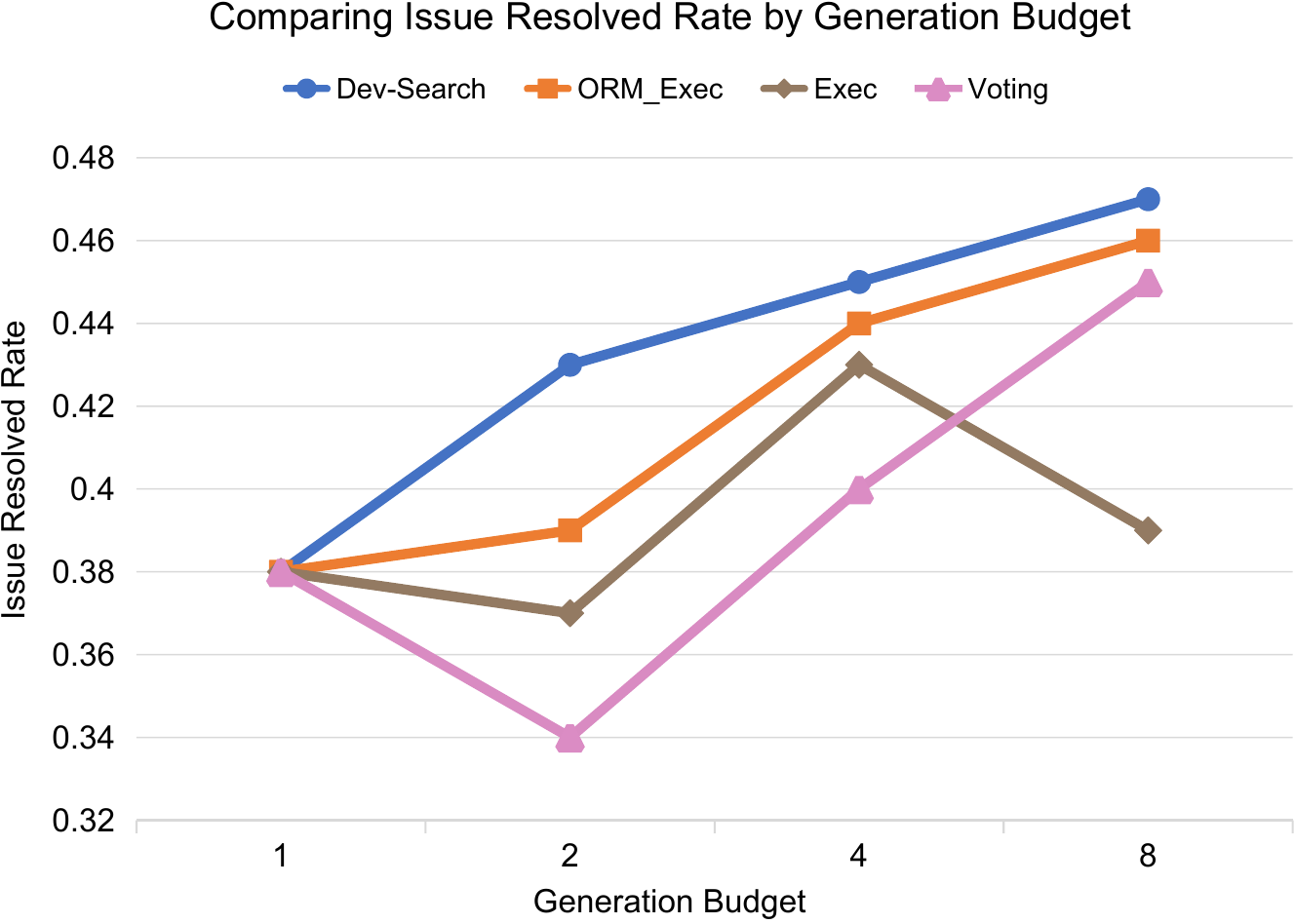}
    % \vspace{-1.0em}
    \caption{The comparative issue-resolution rates under various inference budgets (1, 2, 4, and 8 rollouts).}
    % \vspace{-1.0em}
    \label{fig:prm_search}
\end{figure}

Figure~\ref{fig:prm_search} illustrates the comparative issue-resolution rates under various inference budgets (1, 2, 4, and 8 rollouts). We observe several key findings. First, our proposed Dev-Search strategy consistently achieves the highest resolution rate across all budget conditions, clearly demonstrating its overall effectiveness. Moreover, a distinct test-time compute scaling phenomenon emerges, evidenced by steadily improving performance as inference budgets increase. Conversely, the Exec baseline exhibited an unexpected drop in performance at the highest budget (budget=8). A potential explanation for this performance decline is that execution-based verification alone (specifically, reproducing code functionality) might occasionally yield false positives due to limited coverage and incomplete reproducibility, leading to instability when randomly selecting among candidate patches. Importantly, incorporating the ORM-based tie-breaking method in the ORM\_Exec variant mitigates this issue, achieving stable improvements with increased budgets.

\begin{figure}
    \centering
    \includegraphics[scale=0.42]{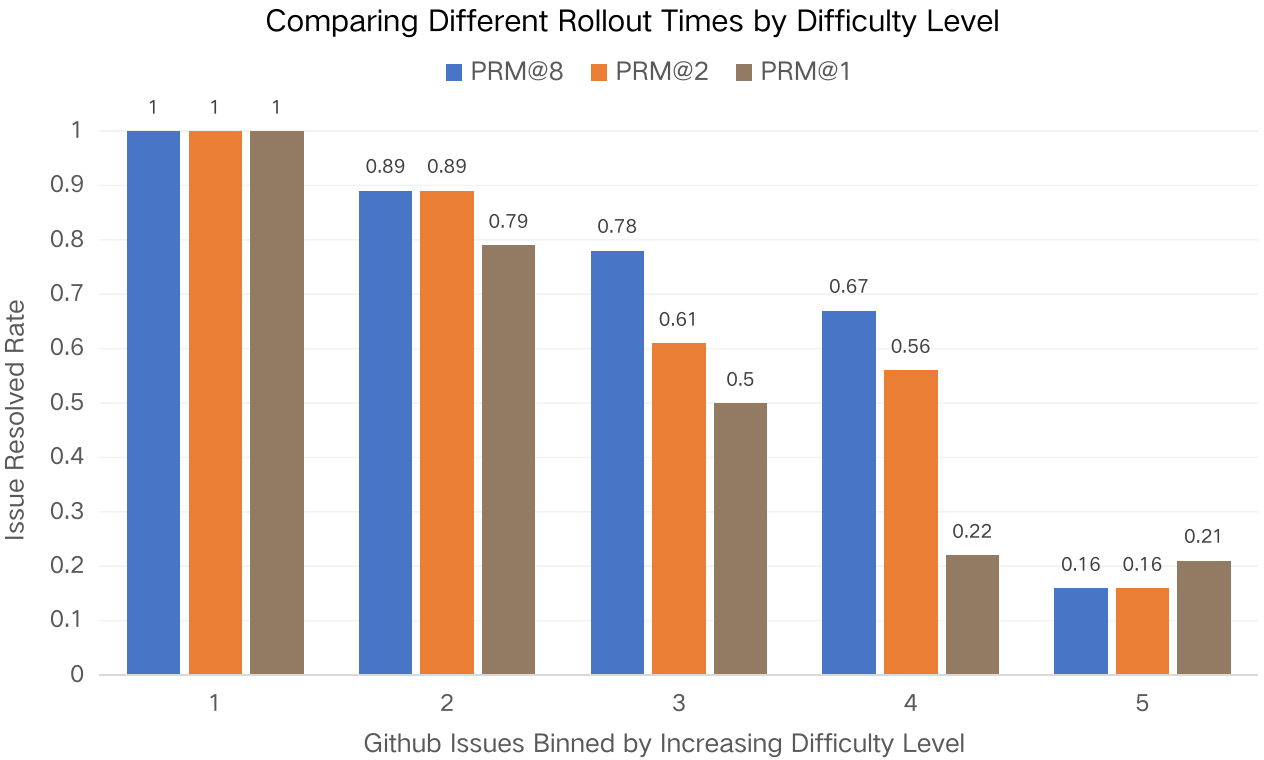}
    % \vspace{-1.0em}
    \caption{Comparison of Issue Resolution Rates: By Difficulty Level and Rollout Time.}
    % \vspace{-1.0em}
    \label{fig:rolloutnums}
\end{figure}

\textbf{Influence of Generation Budgets Across Difficulty Levels.}
Our second experiment analyzed how varying generation budgets impacted agent performance across different issue difficulty buckets. The issues were categorized into five difficulty levels based on their resolution frequency among existing top-ranked agent submissions from the SWE-bench Verified leaderboard (bucket 1: easiest, solved by 25–30 agents; bucket 5: hardest, solved by 5–10 agents). Figure~\ref{fig:rolloutnums} summarizes these results.

As expected, increasing inference budgets consistently improved issue-resolution performance for difficulty levels 1 through 4, particularly notable in difficulty levels 3 and 4. This clearly indicates that additional test-time compute can indeed enhance model performance, allowing the agent to explore broader reasoning trajectories and effectively handle moderately challenging problems.
However, at the highest difficulty level (bucket 5), we observed a slight reduction in resolution accuracy when using higher inference budgets. This counterintuitive finding suggests that, for extremely challenging tasks, the effectiveness of external compute strategies may reach inherent limitations imposed by the model's reasoning capabilities. In other words, beyond certain complexity thresholds, merely allocating more computational budget to external search may offer limited gains without commensurate improvements in underlying model reasoning abilities. Future work should explore combining external compute strategies with complementary internal training improvements to further extend effectiveness on highly challenging tasks.

\section{Related Works}

\subsection{LLM-based Software Engineering Agents}

Generative models have exhibited significant capabilities in code generation. These models have substantially impacted various aspects of software engineering, enabling tasks such as code generation~\cite{ma2023training, pan2024codev, jiang2023automatic, zhu2024domaineval, xu2024cruxeval, tian2024codehalu, zhao2024codejudge}, test generation~\cite{liu2024your, xue2024selfpico, xia2024fuzz4all, li2024dllens}, and code editing and refactoring~\cite{li2023codeeditor, chakraborty2021multi, shirafuji2023refactoring, alomar2024refactor, zhang2024lpr, zhang2023lampr}.
In recent years, AI agents have significantly advanced ASE. These agents enhance project-level SE tasks by integrating diverse capabilities, such as awareness of the running environment~\citep{hong2023metagpt, wang2024codeact, kong2024contrastrepair, xie2024pet}, structured planning and reasoning~\citep{wang2024codeact, cognitionai2023devin, luo2024repoagent}, and leveraging external tools~\citep{zhang2024codeagent, lee2024unified, xue2023acwrecommender, ma2023mulcs, huang2023towards}. Devin~\citep{cognitionai2023devin} notably introduced a milestone end-to-end ASE framework, capable of autonomously planning requirements, utilizing tools such as code editors, terminals, and search engines, and ultimately generating functional code to fulfill user specifications. Its promising capabilities have sparked significant attention within the SE community, inspiring subsequent works, such as SWE-Agent~\citep{yang2024sweagent}, AutoCodeRover~\citep{autocoderover}, and RepoUnderstander~\citep{ma2024understand}. Recently, SWE-SynInfer~\citep{ma2024lingmaswegpt} provided an effective open-source framework to systematically handle software issues, decomposing issue resolution into stages of repository understanding, fault localization, and patch generation. Building upon this framework, we propose \textit{SWE-SynInfer+}, an enhanced version introducing an explicit patch verification phase, where reproduction code is generated to automatically verify and iteratively refine candidate solutions. Besides, a major limitation across existing ASE agents remains their heavy reliance on larger models, which restricts accessibility in real-world deployments. Our work directly addresses this limitation by proposing a scalable inference-time compute framework, explicitly designed to strengthen open-source ASE agents through enhanced reasoning depth and systematic exploration of candidate solutions.

\subsection{Training Software Agents}

Recent advancements have demonstrated the significant potential of leveraging LLMs to tackle complex SE tasks. Existing approaches rely predominantly on proprietary or resource-intensive models such as OpenAI o1~\citep{o1systemcard} or DeepSeek R1~\citep{guo2025deepseek}, achieving strong results but facing barriers related to model accessibility, data transparency, and deployment costs. Efforts have begun to develop open-source alternatives explicitly tailored for emerging SWE tasks. For instance, Lingma-SWEGPT~\citep{ma2024lingmaswegpt} proposes iterative, development-process-centric methods and introduces open model variants derived from Qwen2.5~\citep{hui2024qwen2.5-coder}, achieving improved performance on SWE-bench. SWE-Gym~\citep{pan2024swegym} further advances open-source SWE agent training by providing an environment designed to enhance the Qwen2.5-Coder series (7B and 32B) on SWE-bench tasks. Similarly, SWE-Fixer~\citep{xie2025swefixer} fine-tunes Qwen2.5 models into specialized retrievers and editors for more efficient issue resolution. SWE-RL~\citep{swerl} uses reinforcement learning to improve the Llama model and achieve better issue resolution. In contrast, our work proposes a distinct approach focused explicitly on scalable inference-time compute (TTC) rather than merely scaling model size. Our framework achieves superior or comparable performance to existing state-of-the-art models, while significantly reducing computational demands and resource constraints.

\subsection{Scaling Test-Time Compute} \label{evaluation_related_work}

% Benefiting from the strong general capability of LLMs, LLM-based software engineering agents can handle many important SE tasks, e.g., code generation~\citep{wang2024codeact} and code debugging~\citep{hong2023metagpt}. More recently, SWE-bench team\citep{jimenez2023swe, yang2024sweagent} develop a unified dataset named SWE-bench to evaluate the capability of the agent system to resolve real-world GitHub issues automatically. Specifically, it collects the task instances from real-world GitHub issues from twelve repositories. Consistent with previous evaluation methods, SWE-bench is based on the automatic execution of the unit tests. Differently, the presented test set is challenging and requires the agents to have multiple capabilities simultaneously, including repository navigation, fault locating, debugging, code generation and program repairing, so as to solve a given issue end-to-end. Besides, SWE-bench Verified~\cite{swebenchverified} is subset of SWE-bench, and it have a similar diversity and distribution of repositories as the original version. Due to the smaller test cost and more detailed human filtering, SWE-bench Verified is officially recommended as the benchmark of LLM-based SE agents. Therefore, consistent with previous methods~\citep{yang2024sweagent,autocoderover,xia2024agentless}, we report our performance on SWE-bench Verified.

Recent advancements in software engineering agents, leverage external tools like parallel trajectory generation, voting mechanisms~\citep{xia2024agentless}, and execution verification~\citep{swerl, ehrlich2025codemonkeys} to enhance solution quality. For example, SWE-Gym~\citep{pan2024swegym} trains an ORM to select the highest-scoring trajectory from parallel generations, while Agentless~\citep{xia2024agentless} employs a voting mechanism to normalize and rank candidate patches, choosing the most frequent one. Although effective, these methods do not address intermediate steps in the workflow. Extensions like CodeMonkeys~\citep{ehrlich2025codemonkeys} and SWE-RL~\citep{swerl} generate reproduction code to validate patch correctness, offering functional feedback. Similarly, Nebius~\citep{nebius} introduces PRMs to guide action selection. Yet, these frameworks still emphasize either trivial or final actions rather than systematically addressing all critical stages of development. Moreover, they fail to explore the potential of internal TTC to dynamically scale reasoning capabilities. Our work bridges these gaps by proposing a novel framework that integrates targeted search at three pivotal development phases alongside Long CoT training. To the best of our knowledge, this is the first empirical demonstration of test-time scaling within software engineering agents.

\section{Limitation and Threats to Validity}

While SWE-Reasoner 32B (TTC) demonstrates promising results in automated software improvement, it is important to acknowledge several limitations that affect both the current approach and the broader generalizability of our findings:
\textbf{Inference Efficiency.}
Although test-time compute scaling substantially improves model performance, it can also degrade inference efficiency, particularly for interactive tasks such as real-time code completion or conversational code assistance. In contrast, for end-to-end software issue resolution where short delays are acceptable, test-time scaling remains practical. We also observe that SWE-Reasoner and OpenAI-o1 partially adapt their reasoning depth to problem complexity, suggesting that future work could explore more fully adaptive inference-time mechanisms—automatically adjusting the extent of reasoning based on task difficulty or runtime constraints.
\textbf{Automated Solution Verification.}
Despite strong results on the SWE-bench dataset, our training data remains relatively small, primarily due to the constraints of verifying solutions in real-world software environments. Few datasets capture the entire testing and debugging lifecycle, and automatically setting up complex project environments with myriad dependencies is a significant challenge for current tooling, which frequently has a low success rate. Future research could improve the end-to-end capabilities of SWE agents by developing more precise, large-scale automated environment-setup frameworks. Integrating these frameworks with reinforcement learning or other adaptive training methods might further enhance the robustness and applicability of automated software engineering systems. Despite these limitations, SWE-Reasoner 32B (TTC) constitutes a significant step forward in automated software engineering. The challenges outlined above highlight opportunities for continued investigation and improvement. We plan to leverage these insights to evolve SWE-Reasoner into a more robust, adaptable, and effective solution, ultimately aiming to assist developers across the full spectrum of software development tasks.

\section{Conclusion}

In this work, we introduced a unified Test-Time Compute (TTC) scaling framework to enhance the code reasoning capabilities of software engineering agents using personally deployable open-source LLMs. Internally, we proposed a development contextualized trajectory synthesis method, leveraging realistic multi-stage reasoning trajectories extracted from high-quality GitHub repositories. This method, combined with repository-aware rejection sampling, significantly improves the model’s internal reasoning capabilities. Externally, we developed a development-process-based search strategy that focuses computational resources at critical decision-making points, utilizing specialized reward models and execution verification to efficiently prune less promising trajectories. Evaluations conducted on the challenging SWE-bench Verified demonstrate that our 32B model, with targeted inference-time scaling, achieves a state-of-the-art 46\% issue resolution rate, outperforming larger models. Additionally, we provided the first empirical validation of the test-time scaling phenomenon within SWE agents, revealing effective dynamic allocation of computational resources to address increasingly complex software engineering tasks. Future work includes extending our unified TTC framework to broader software engineering tasks, exploring adaptive computation allocation strategies informed by task difficulty prediction, and investigating TTC's applicability across different software engineering environments and domains.

\begin{acks}
We would like to express our gratitude to Wenhao Zhang\footnote{\url{https://doc.agentscope.io/tutorial/swe.html}} and Zhipeng Xue\footnote{\url{https://zhipengxue97.github.io/}} for their invaluable feedback and suggestions on the manuscript.
\end{acks}

\bibliographystyle{ACM-Reference-Format}
\bibliography{sample-base}

% \appendix









\end{document}